\documentclass{ws-mplb}
\usepackage{amsfonts,amssymb,amsmath}
\usepackage{graphicx}
\usepackage[super,sort,compress]{cite} 

\begin{document}

\markboth{Hrishit Banerjee}{Understanding the role of correlations in complex oxides under strain and oxide heterostructures}

%
\catchline{}{}{}{}{}
%

\title{Understanding the role of exchange and correlations in complex oxides under strain and oxide heterostructures}

\author{\footnotesize Hrishit Banerjee}

\address{Institute of Theoretical and Computational Physics, Graz University of Technology,\\ NAWI Graz, Petersga{\ss}e 16\\ Graz, 8010, Austria. \\
h.banerjee10@gmail.com}

\maketitle


\begin{abstract}
The study of complex oxides and oxide heterostructures have dominated the field of experimental and theoretical condensed matter research for the better part of the last few decades. Powerful experimental techniques like molecular beam epitaxy and pulsed laser deposition have made fabrication of oxide heterostructures with atomically sharp interfaces possible, while more and more sophisticated handling of exchange and correlations within first principles methods including density functional theory (DFT) supplemented with Hubbard U corrections and hybrid functionals, and beyond DFT techniques like dynamical mean field theory (DMFT) have made understanding of such correlated oxides and oxide interfaces easier. The emergence of the high mobility two dimensional electron gas with fascinating properties like giant photoconductance, large negative magnetoresistance, superconductivity, ferromagnetism, and the mysterious coexistence of the latter two have indeed caught the attention of condensed matter community at large. Similarly strain tuning of oxides have generated considerable interest particularly after the recent discovery of piezoelectric methods of strain generation. Theoretical understanding and prediction of the possible exotic phases emerging in such complex oxides both under strain and in heterostructures will eventually lead to better design of device applications in this new emerging field of oxide electronics, along with possible discovery of exotic physics in condensed matter systems which may be of wider significance! In this review we briefly look at theoretical studies of novel phenomena in oxides under strain and oxide heterostructure, and try to understand the role of exchange and particularly correlation in giving rise to such exotic electronic states. This review though primarily focuses on the theoretical aspects of understanding the different mechanism of the phenomena of emergence of exotic phases, does provide a unique overview of the experimental literature as well, accompanied by the theoretical understanding such that relevant device applications can be envisaged.
\end{abstract}


\section{Introduction}
Complex oxides represent a class of materials with a plethora of fascinating physical
properties, under various conditions. The intriguing interplay of charge, spin, and orbital ordering in these systems coupled with lattice effects opens up a scientifically rewarding playground for both fundamental and application-oriented research. In particular transition metal oxides (TMO) continue to attract a great deal of attention both experimentally and theoretically spanning over many decades, due to their several interesting properties, e.g. Mott transition,
unconventional superconductivity, ferromagnetism, antiferromagnetism, low-spin/high-spin transition, ferroelectricity, antiferroelectricity, colossal magnetoresistance, charge ordering etc.

Metal oxide perovskites constitute one of the largest class of oxide materials. The structure of these perovskite oxides are mainly derived from cubic structure. However many compounds at low temperatures often exist with lower symmetry structures such as in tetragonal, orthorhombic or hexagonal symmetries. 
However often these systems show various structural distortion among which most common are gadolinium ferrite distortions involving rotation and tilt of the metal-oxygen octahedra and Jahn-Teller distortions involving distortion in metal-oxygen bond lengths, both of which lead to the splitting of commonly observed $t_{2g}$ and $e_g$ orbitals into further non-degenerate states. Application of strain often plays a significant role in correlated oxides in modifying the electronic states of the material. It has been seen that correlation itself may be tuned by application of strain in certain oxides. \cite{strain-oxide, kim}

Developments in recent experimental methods like molecular beam epitaxy (MBE) or pulsed laser deposition (PLD) have lead to the possibility of layer by layer growth of perovskite oxides on one another. Such layered structures with more than one oxide compound in the structure are called oxide heterostructures.
The pioneering work by Ohtomo \& Hwang \cite{hwang-ch1} established that a two dimensional electron gas (2DEG) of high carrier density and high mobility  
is formed at the interface (IF) of two perovskite oxides, like LaAlO$_3$(LAO) and SrTiO$_3$(STO), both of which are band insulators, and \cite{manhart-ch1} neither of which share the
properties seen at the interface. Interfaces can exhibit electrical conductivity\cite{hwang-ch1}, superconductivity\cite{gariglio-ch1}, ferromagnetism \cite{bert-ch1}, large negative in-plane magnetoresistance\cite{ben-ch1}, and giant persistent photoconductivity\cite{tebano-ch1}. The most intriguing property \cite{brinkman-ch1} among those shown by interfaces seem to be the puzzling existence of both ferromagnetism and superconductivity! 

Under the right conditions, the LaAlO$_3$/SrTiO$_3$ interface is electrically conductive, like a metal. The angular dependence of Shubnikov-de Haas oscillations indicate that the conductivity is two-dimensional, leading to its reference as a two-dimensional electron gas (2DEG)\cite{caviglia-ch1}.
The most prevalent hypothesis for conductivity of the 2DEG is the Polar Catastroph\'{e} Hypothesis, which postulates that LaAlO$_3$, which is polar in the (001) direction (with alternating sheets of positive and negative charges), act as an electrostatic gate on the insulating SrTiO$_3$. It is an allusion to the  scenario where a sudden jump in potential occurs from charge neutral SrO and TiO$_2$ layers to the charged LaO and AlO$_2$ layers which have charges of +1 and -1 respectively. Due to this jump in charge there is a corresponding jump in electric field and the voltage in the LaAlO$_3$ section builds up forever and diverges eventually. Of course it is highly unlikely that such a divergence in potential can exist in nature and hence it is avoided by means of an electronic reconstruction, where, half an electron per unit cell is transferred to the interface TiO$_2$ layer and consequently half a hole is transferred to the surface AlO$_2$ layer. This explains the origin of half an electron charge at the interface of two insulators. 

The study of complex oxides and oxide heterostructures with their plethora of application possibilities and exotic states have been the holy grail of the upcoming field of oxide electronics.
There is an ongoing exchange between theory and corresponding experiments in this exciting new field of 2D electron gases, and its several application possibilities. A deeper understanding of all the intrinsic novel properties is therefore needed, which requires a substantial scientific effort to explore extensively, and in this context the atomistic insight on these materials is extremely relevant at this point of time. Much of the understanding in this field has been through electronic structure calculations although much is left to be explored. The use of electronic structure calculations to understand such systems is a very practical approach as it has the potential to unveil undiscovered phenomena and properties in systems based on  crystal structure of materials that are available in the process of experimental synthesis. By performing electronic structure calculations, we always include the  structural and chemical aspects of the system in full rigour without loss of generality. First principles calculations thus form the backbone for obtaining theoretical understanding of oxides. This review however not only focuses on the theoretical aspects of understanding the different mechanisms involved in the emergence of exotic phases in oxides, but also provides a unique overview of the experimental literature as well, accompanied by the theoretical understanding such that relevant device applications can be envisaged.

In this brief review we re-visit several studies on both complex oxides under strain with a rich phase diagram and which have shown a very interesting correspondence between exchange-correlations and strain effects, and also important aspects of oxide heterostructures focusing particularly on the importance of strong correlations in giving rise to novel states in such systems. First we highlight in brief the theoretical methods which are particularly tailored to studying the effect of exchange and correlation in oxides. In the subsequent sections we describe the electronic structure of several oxides under strain and oxide heterostructures where exchange and correlation has been seen to play a very crucial role in giving rise to exotic electronic states. Finally we suggest further future studies which may be carried out from a theoretical perspective to better understand the role of electron exchange and correlation in oxides, which is expected to lead to better control and design of device applications. A thorough and concise understanding of the effect of correlation and exchange in oxide systems is imperative for such applications.

\section{Theoretical methodology to study the effect of exchange and correlation}
Density functional theory has been very successful in explaining and predicting the electronic structure of materials for several decades now. However the shortcoming of LDA, or GGA methods is the lack of proper implementation of the exchange-correlation functional. This is of paramount importance in strongly correlated materials, and a static implementation of the ad-hoc energy cost in the form of Hubbard $U$ is not enough in a lot of cases. Therefore one might have to look at better approximations for exchange and correlation particularly in case of oxides which are in general strongly correlated materials.

In the following we briefly describe the two most popular methods of correction for exchange and correlation - (i) The inclusion of Hartree Fock exact exchange in DFT in the form of a hybrid functional within DFT and (ii) Dynamical Mean Field Theory. 

\subsection{Hybrid functionals}
Hybrid functionals are approximations to the exchange-correlation energy functional in DFT that includes a portion of exact exchange from Hartree-Fock theory with the rest of the exchange-correlation energy from DFT functionals. The exact exchange energy functional is expressed in terms of the Kohn–Sham orbitals rather than the density, so is termed an implicit density functional. Several hybrid functionals like B3LYP (Becke, 3-parameter, Lee-Yang-Par), PBE0 (Perdew, Berke, Ernzerhof with 0.25\% exact exchange), HSE (Heyd–Scuseria–Ernzerhof) etc are available among which the most popular and widely used form is the HSE hybrid functional.

The HSE  exchange–correlation functional \cite{hse} uses an error-function-screened Coulomb potential to calculate the exchange portion of the energy in order to improve computational efficiency, especially for metallic systems: the functional used in HSE calculation can be mathematically expressed as,

\begin{align}
E_{XC}^{HSE}(\omega)= & \alpha E_X^{HF,SR}(\omega)+(1-\alpha)E_X^{PBE,SR}(\omega) \notag\\ & + E_X^{PBE,LR}(\omega)+E_C^{PBE}
\end{align}
where $\alpha$ is the mixing parameter and $\omega$ is an adjustable parameter controlling the short-rangeness of the interaction. Here $E_X^{HF,SR}$ denotes the short range HF exchange functional, $E_X^{PBE,SR}$ denotes the short range PBE exchange functional, $E_X^{PBE,LR}$ indicates the long range PBE exchange functional, and $E_C^{PBE}$ refers to the correlation functional as given by PBE. The standard value of $\omega$=0.2 (referred to as HSE06) along with varying values of $\alpha$  of 0.15, 0.20, 0.25, and 0.30 are generally used in calculations.

\subsection{Dynamical Mean field Theory}
The main short coming of DFT is the treatment of electron-electron correlation, which is treated in average manner in DFT. Although such a prescription is highly successful for most of the cases, but it fails when  electron-electron correlation become strong as for $\textit{d}$ and $\textit{f}$ electron systems. Therefore the  most obvious choice to deal with this problem is to couple the LDA low energy Hamiltonian with the missing correlation part, which defines many body Hamiltonian, such as Hubbard model \cite{DMFT1}, Heisenberg Model, t-J 
model etc but now derived in an ab-initio way. The constructed model then can be solved with $\textit{Dynamical Mean Field Theory}$ (DMFT) \cite{DMFT2}. DMFT is non-perturbative method to take care of the electronic interaction beyond the independent electron approximation. It maps the non-tractable many body lattice model to the local quantum impurity model which can be solved by Quantum Monte Carlo, Exact Diagonalization or other methods.


The Hubbard model is considered as the original lattice model. To construct the single-site DMFT, one can assume that the self-energy be local.

The most general reference system with a local self-energy and arbitrary $\omega$-dependence is given by a single interacting (“impurity”) site with $U\neq 0$ hybridizing with a continuum of non-interacting bath degrees of freedom, i.e., by the single-impurity Anderson model,

\begin{equation}
H'=\sum_{\sigma}\epsilon^{imp}c^{\dagger}_{\sigma}c_{\sigma}+\frac{U}{2}\sum_{\sigma}n_{imp,\sigma}n_{imp,-\sigma}+\sum_{k\sigma}\epsilon_{k}a^{\dagger}_{k\sigma}a_{k\sigma}+\sum_{k\sigma} (V_{k}c^{\dagger}_{\sigma}a_{k\sigma}+h.c.)
\end{equation}

The effective self consistent quantum impurity model, deals with a set of local quantum mechanical degrees of freedom that interacts with a bath created by all other degrees of freedom on other sites. This approximation is exact in infinite dimensions. 

The local Green function on the impurity site is
\begin{equation}
     G^{(imp)}(\omega)=\frac{1}{\omega+\mu-\epsilon^{imp} -\Delta(\omega)-\Sigma'(\omega)}
\end{equation}

The bath parameters, namely the hybridization strengths $V_k$ and the on-site energies $\epsilon_k$, enter the formalism via the hybridization function only:

\begin{equation}
   \Delta(\omega)=\sum_{k\sigma}\frac{V_{k}^2}{\omega+\mu-\epsilon_{k}}
\end{equation}

We will use the self-energy of the reference system as an approximation for the lattice model,
$\Sigma'(\omega)=\Sigma(\omega)$. This implies that, 

\begin{equation}
G^{(imp)}(\omega)=G^{(loc)}(\omega)
\end{equation}
which is the well known self-consistency condition of DMFT.

In the combined LDA+DMFT approach the one particle input of DMFT Hamiltonian comes from LDA band structure calculations, which is an useful methodology to take care of material specific aspect. The corresponding local Green's function is calculated via the Dyson equation, which has the form,
\begin{equation}
G(\omega_{n}) = \sum_{k} \left[ (\omega_{n} + \mu ) I - H^{LDA}(k) - \Sigma(\omega_{n}) \right] ^ {-1},
\label{DMFT1}
\end{equation}


where, $H^{LDA}(k)$ is the LDA few band Hamiltonian defined in Wannier function basis. $\mu$ is the chemical potential calculated self consistently via total number of electrons, $\omega_{n}$ are the Matsubara frequencies related to the temperature ($\beta=1/k_BT$) via the relation $ \omega_{n} = \frac{(2n+1)\pi}{\beta}$. $\Sigma$is the self energy matrix related to local and bath green's function via the following relation;

\begin{equation}
\Sigma =g^{-1} - G^{-1}
\label{DMFT2}
\end{equation}
The local Green's function therefore has to be calculated self consistently with the condition that implies the local Green's function to be the same as the corresponding solution of the quantum impurity problem.

\section{Complex oxides under strain}
Understanding the methods of calculation, we now turn our attention to the various studies on correlated oxides under strain. Strain tuning has been found to be particularly useful in bending properties to suit the needs of device applications. Strain has been shown to be effective in tuning both material properties of use in practical applications and also in tuning electronic correlations and magnetism which is interesting from a theoretical standpoint thereof. 

Hwang et al show that when complex oxides are grown as thin films, their chemical and physical properties can be modified to be markedly different from their bulk form, providing additional degrees of freedom in materials design. They explored the landscape of strain-induced design of complex oxides in the context of oxygen electrocatalysis and ferroelectricity. \cite{strain-oxide} Strain driven modification of metal-oxygen bond length and octahedral distortion in perovskites has been shown to influence oxide electronic properties. Influence of strain on material properties relevant for oxygen electrocatalysis and ferroelectricity has been studied along with the advances in state-of-the-art thin-film fabrication and characterization that have enabled a high degree of experimental control in realizing strain effects in oxide thin-film systems. In oxygen electrocatalysis, leveraging strain has not only resulted in activity enhancements relative to bulk unstrained material systems but also revealed mechanistic influences of oxide phenomena, such as bulk defect chemistry and transfer kinetics, on electrochemical processes. Similarly for ferroelectric properties, strain engineering can both enhance polarization in known ferroelectrics and induce ferroelectricity in material systems that would be otherwise non-ferroelectric in bulk.

Franchini et al \cite{kim} showed that epitaxial strain offers an effective route to tune the physical parameters in transition metal oxides. The effects of strain on the bandwidths and crystal field splitting has long since been known, however recent experimental and theoretical works have shown that also the effective Coulomb interaction changes upon structural modifications. This particular effect has huge influence in current materials design studies based on epitaxy-based material synthesis. In their theoretical study they show that the response upon strain is strongly dependent on the material for prototypical oxides with a different occupation of the $d$ shell. For 3 oxides, LaTiO$_3(d^1)$, LaVO$_3(d^2)$, and LaCrO$_3(d^3)$, from a theoretical standpoint based on constrained random phase approximation calculations they systematically study the evolution of effective Coulomb interactions (Hubbard U and Hund's J) with the application of epitaxial strain. This effectively proves that correlations are significantly altered by the effect of strain on oxides, which is thought to have far reaching effects on emergence of exotic properties.

\begin{figure}[h]
	\centering
	\includegraphics[width=\columnwidth]{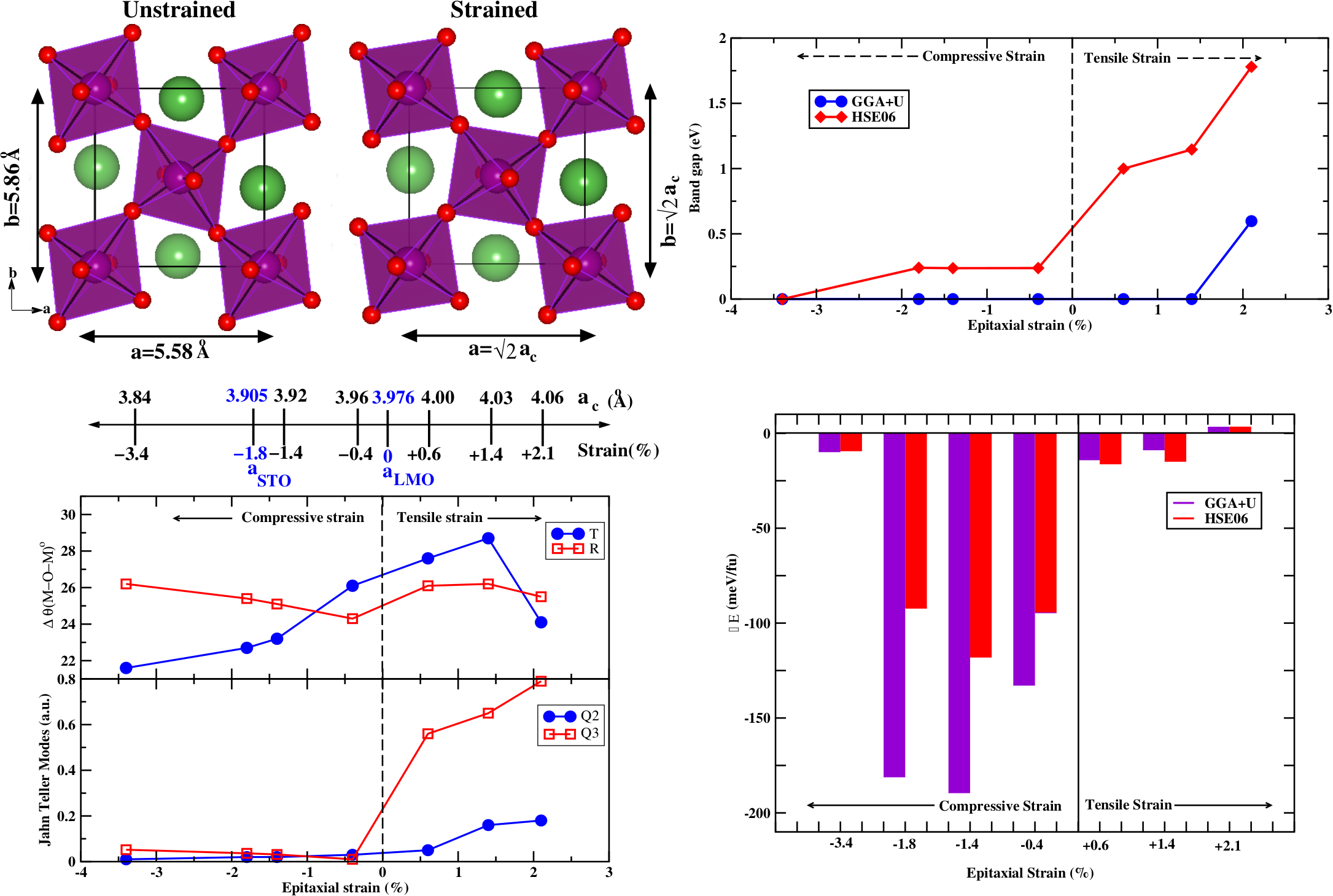}
	\caption{Figure showing the structures of unstrained and epitaxially strained LMO and respective energies on application of strain. The top left panel shows how the LMO is strained to a square lattice. The middle left panel shows the different epitaxial strains and corresponding values of the $a$ lattice parameter. The bottom left panel shows the rotation and tilt in the Mn-O octahedra and the corresponding Jahn Teller distortions at various values of strains. The top right panel shows the band gap in LMO as a function of strain for GGA+$U$ and HSE06 calculations.
		The bottom right panel shows energy differences $\Delta E=E_{FM}-E_{AFM}$ between FM and A-AFM configurations at various square matching epitaxial strain values for both GGA+$U$ and HSE06, showing that the trend in magnetic ground state for both GGA+$U$ and HSE06 are the same. Figure adapted from Banerjee et al \cite{banerjee-hse}}
\end{figure}

In our previous study \cite{banerjee-hse} we have already shown the effect of epitaxial strain on LaMnO$_3$ which is a Jahn Teller driven correlated oxide. We have shown that compressive epitaxial strain exerted by epitaxial matching of LMO to a square substrate significantly reduces the Jahn Teller distortion in LMO which in turn favours a ferromagnetic (FM) ground state instead of the antiferromagnetic state in bulk LMO. Application of tensile strain beyond a certain critical value favours antiferromagnetism. This is demonstrated in Fig. 1. It has also been shown that the ferromagnetic insulating phase in orthorhombic LMO strained to an STO substrate arises from the charge disproportionation as obtained within the formulation of hybrid functionals, as demonstrated in Fig. 2. This is also the case with other moderate values of compressive strain where charge ordering gives rise to FM insulating behaviour over a wide range of compressive strain values. For tensile strain values we find FM-I state to be driven by an alternate orbital ordering, as shown in Fig. 2. For large values of tensile strain an AFM-I phase is stabilised which is also the ground state for bulk unstrained LMO where the LMO in plane lattice parameters are not fitted to that of a square lattice. Thus we find that not only strain but also matching the lattice parameters to a square lattice plays an important role in this case. This also demonstrates that simple GGA+$U$ is unable to capture the effects of electron exchange and hence does not correctly capture the ground state properties of correlated oxides.
In a recent study \cite{banerjee-dmft} we also show that this transition may also be driven by strong correlations from a DMFT based many body perspective. For the ``strained-bulk'' structure, corresponding to a compressive strain of 1.8\%, it was found that DFT+DMFT yields a ferromagnetic insulating solution for small enough temperature. The critical temperature $T_\text{C}$ is found to be $\sim$ 100K. Thus we can also confirm the higher energy state to be a ferromagnetic insulator from the perspective of ab initio DMFT calculations, which is not possible within the regime of DFT calculations even with the inclusion of exact exchange. Our studies show from the perspective of correlation and dynamics, the emergence of the exotic ferromagnetic insulating state, which has been touted to be immensely important in case of many spintronics based device applications. It also stresses on the fact that strain tuning of various exotic electronic phases is possible, and the effect of correlations are deeply entwined in such processes.

\begin{figure}
	\centering
	\includegraphics[width=\columnwidth]{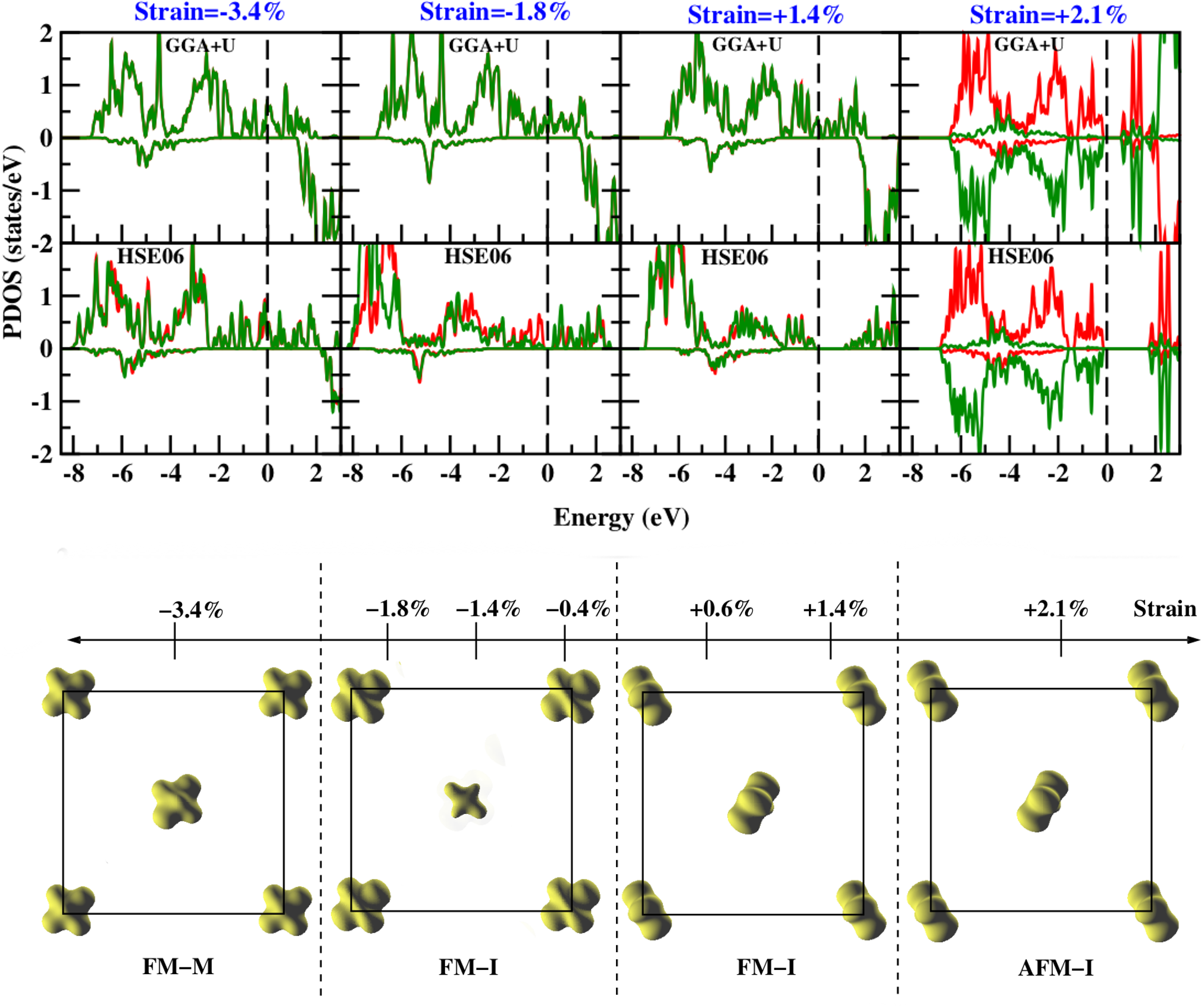}
	\caption{Figure showing the variation of electronic structure properties with strain applied on LMO. Top panel shows the projected DOS on the two equivalent Mn atoms in the unit cell for both GGA+U and HSE functionals, for different values of compressive and tensile strain. This shows that HSE06 better captures the effect of exchange than GGA+U. The lower panel shows the variation in charge density in the unit cell with the same range of strain values.}
\end{figure}

Ruthenates and Iridates have both strong influences of correlation and in case if Iridates also the presence of strong spin-orbit coupling accounts for very exotic magnetic and electronic states. BaRuO$_3$ \cite{bro} has already shown presence of various different polytypes at various applied pressures experimentally, and thus shows structural phase transitions from rhombohedral, to hexagonal and finally to cubic at high pressures of 15GPa. Theoretical studies on the cubic phase shows a transition from a generalised fermi liquid to a correlated non-fermi liquid behaviour. This of course indicates the importance of strong correlations in ruthenates. 
Considering SrRuO$_3$ (SRO) it has been shown experimentally that SRO films undergo a structural transition from the low-temperature orthorhombic phase to the high-temperature pseudocubic phase under the influence of epitaxial strain \cite{sro-strain1}. It has also been shown that the ferromagnetic transition temperature T$_C$ for the compressively strained SRO is as high as 155 K, while T$_C$ of the film under tensile strain is only 100 K \cite{sro-strain2}. Thus even the type of applied strain modifies the material properties. Longitudinal resistivity of compressively strained SRO was found to be lower than that of SRO under the tensile strain due to the enhanced mobility in case of compressive strain. Magnetic anisotropy also exhibited strong dependence on the substrate-induced epitaxial strain. The film under compressive strain had  uniaxial magnetic easy axis along the out-of-plane direction, while the easy axis of the film under tensile strain was found to be along the in-plane direction. 
SrIrO$_3$ is a strong spin orbit coupled weak correlated metal which in contrast to bulk, shows insulating behavior in strained films \cite{sio-strain1}. The metal insulator transition has been shown to be either of Anderson or unconventional Mott-Anderson type depending on the methodology of strain generation either by reducing thickness (on best lattice matched substrate) or changing degree of lattice strain (by lattice mismatch between film and substrates) on films, respectively \cite{sio-strain2}. In the correlated metallic phases of SrIrO$_3$, like BRO, a non-Fermi liquid behaviour emerges in resistivity measurements. Transport measurements under magnetic fields show negative magneto-resistance at low temperature for thin-films under compressive strain.

There may be several ways of generating strain. The most common is applying a biaxial strain by growing one lattice on another epitaxially in the form of a heterostructure through methods like MBE or PLD. This of course in case of oxide heterostructures \cite{hwang-ch1}, also generates the polar 2DEG, if accompanied by a potential divergence. However modern piezoelectric methods of strain generation recently developed by Hicks et al \cite{hicks} also enables one to apply uniaxial strain on crystals and films. This particular method is quite intriguing since one may study the effects of strain without dealing with the polar 2DEG right away, and also study the effect of straining one axis in respect of another in case of non cubic geometries without dealing with the extra complications involved with lattice mismatch as in the case of epitaxial growth.

\section{Oxide Heterostructures}

Next we concentrate on oxide heterostructures, grown epitaxialy by MBE or PLD techniques. As mentioned previously these systems have been at the helm of both experimental and theoretical condensed matter research due to the plethora of interesting properties the 2D electron gas shows which arises at the interface due to the electronic reconstruction to avoid divergence catastroph\'{e}. 

The LAO/STO interface in which the 2D electron gas was first seen to emerge has also been the most studied system. In case of LAO/STO carrier densities are experimentally found \cite{siemons-ch1} to be an order of magnitude smaller than (e/2) which is expected out of polar catastrophe model. Moreover the IFs have been reported to be insulating below a critical thickness of LAO layers \cite{thiel-ch1}.

The $n$-type interface in LAO/STO heterostructures, being the conducting interface has been widely studied.\cite{manhart-ch5,huijben-ch5,review1-ch5,review2-ch5} It exhibits gate-tunable 
superconductivity and in addition shows signatures of local moments and possible ferromagnetism\cite{brinkman-ch5,reyren-ch5,bert1-ch5,li-ch5} coexisting with the superconductivity. The density of itinerant carriers, however, is consistently found to be an order of magnitude smaller \cite{siemons-ch5,basletic-ch5,thiel-ch5} than
$0.5e^{-}$/interface, the value expected from the polar catastrophe model. In addition, the interfaces are insulating, rather than being metallic, below a 
certain critical thickness of LAO layers.\cite{thiel-ch5}

A more recent development is the study of the $n$-type interface between the Mott insulator GTO and the band insulator STO grown by molecular beam epitaxy.\cite{moetakef-ch5, cain-ch5, moetakef1-ch5} Remarkably, the GTO/STO samples give rise to 2DEGs with carrier densities of 
$0.5e^{-}$/interface, exactly as expected from the ideal polar catastrophe scenario. Furthermore, the GTO/STO interface is found to be conducting irrespective of layer thickness of GTO, and hence there is no thickness threshold for metallic behavior. In both these respects GTO/STO seems to be qualitatively different from LAO/STO. 

\begin{figure}
	\centering
	\includegraphics[width=\columnwidth]{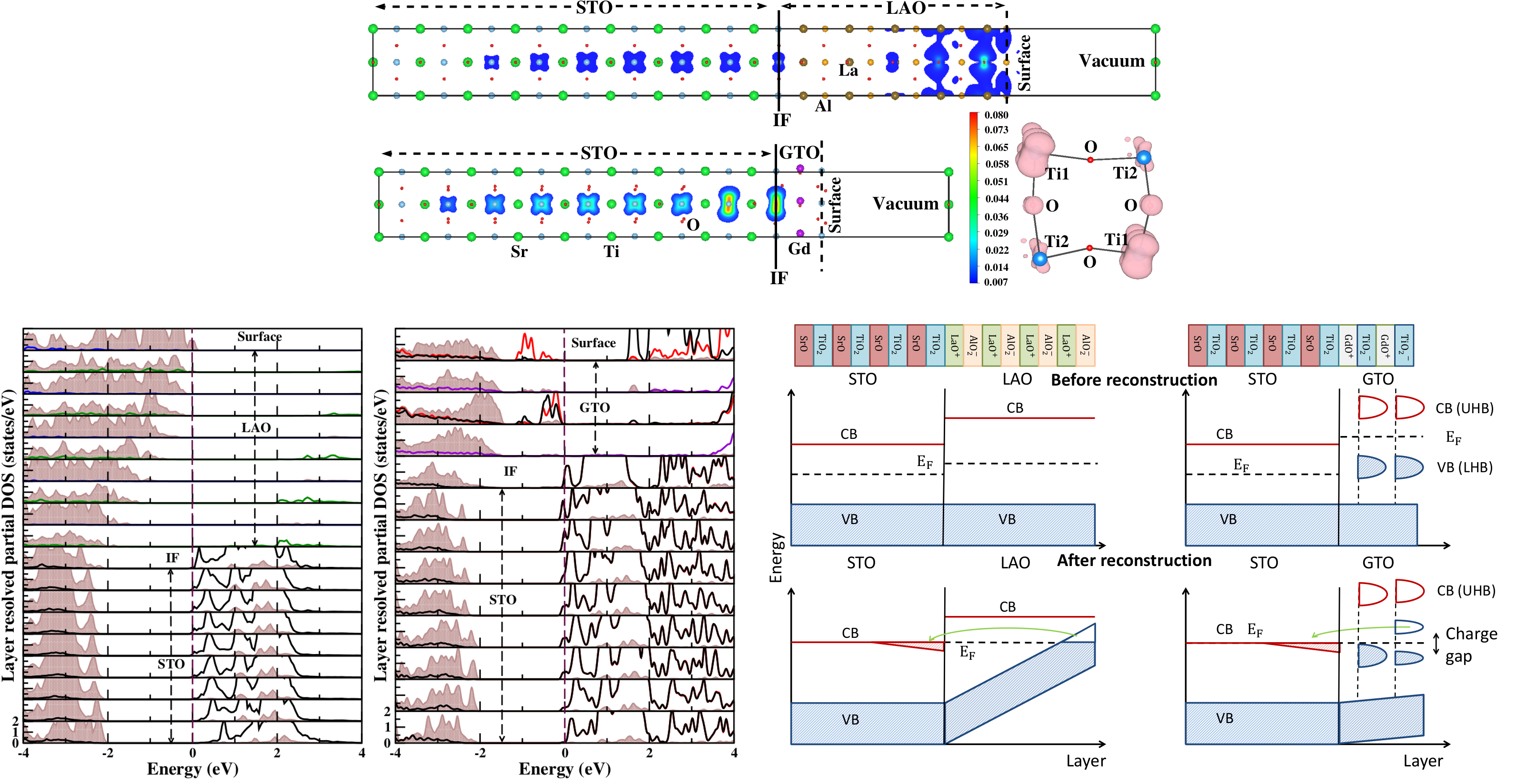}
	\caption{The figure shows the comparison of electronic structure of LAO/STO and GTO/STO. The top panel shows the charge density for both with the charge disproportionation in top layer of GTO/STO, The bottom panel shows the band bending from the DFT perspective and the schematic diagram explaining the difference in critical thickness for conduction for LAO/STO and GTO/STO. Figure adapted from Banerjee et al \cite{banerjee-ch6}}
\end{figure}

The GTO/STO interface also shows many other interesting properties. Quantum oscillation experiments have corroborated the presence 
of confined 2DEG and revealed the possibility of a sub-band structure different from LAO/STO.\cite{moetakef-ch5, cain-ch5 ,moetakef1-ch5} Below a certain STO thickness, the interface has been found to 
exhibit signs of ferromagnetism, which might be an intrinsic property induced by electronic correlations in the high-density 2DEG, rather than induced by the proximity to ferromagnetic GTO layer.\cite{moetakef2-ch5} The signatures of strong electronic correlations have also been observed in transport measurements\cite{moetakef3-ch5} and this has led to the theoretical proposal of dimer Mott insulator for single SrO layer sandwiched between GTO layers.\cite{chen2013-ch5}

While the LAO/STO interface has been thoroughly studied by electronic structure calculations,\cite{lao-elec-ch5,pentcheva-ch5,chen-ch5}
much less was known about GTO/STO. The specific problem of single SrO layer in a GTO matrix in the superlattice geometry has been studied\cite{chen2013-ch5, walle-ch5,lechermann-ch5} by a variety of techniques, first-principles, model Hamiltonian as well as combined density functional theory (DFT) and 
dynamical mean field theory (DMFT).

In a previous study we have carried out a detailed first-principles based investigation of LAO/STO and GTO/STO heterostructures \cite{banerjee-ch6} and found that very different pictures emerge in case of thin film-substrate geometry due to the differences in electronic reconstructions in the two systems. We find a full 0.5 $e^-$/Ti conducting charge at the interface even for 1 u.c. thick GTO on STO substrate, consistent with experiments. On the other hand, in case of LAO on STO, the transferred charge only increase gradually with thickness from a small value of $\sim 0.14 e^-/Ti$ above a critical thickness of about 4. Additionally, we find the fate of the surface layers, which host the neutralizing charges for the interface carriers, to be quite distinct. The electronic states derived from O p orbitals at the surface LAO layer turn out to be metallic within DFT, as shown in Fig. 3, but are experimentally found to be localized, possibly due to disorder or surface imperfections like oxygen vacancies. On the other hand, the Ti $d$ states at the top most layer of GTO/STO correspond to a doped Mott insulating layer of GTO and stays insulating by opening a charge gap via an interesting correlation driven charge disproportionation, as shown in Fig.3. As for the band alignments, also demonstrated in Fig. 3, in case of LAO/STO, the valence band offset is small, and the valence band maxima of LAO and STO which is the upper edge of filled O-p bands, are almost aligned. The experimentally measured bulk band gap of STO is 3.3 eV, while that of LAO is 5.6 eV.  For conduction, therefore a large band bending with VB maxima of LAO aligning with conduction band minima of STO is needed. The necessary band bending is estimated to be about the  same as band gap of STO {\it i.e.}, 3.3 eV. The screened potential shift of a 4-5 u.c. of LAO is just sufficient  to allow the necessary band bending of 3.3 eV. For GTO/STO, on the other hand, the upper edges of VB of GTO and STO are misaligned. In case of GTO, it is the Ti $d$ lower Hubbard band, which energetically lies far above the upper edge of the VB of STO which is the O $p$ band. After electronic reconstruction, in the topmost layer the Ti $d$ LHB splits into two bands, due to charge disproportionation, requiring a small bending of $\approx$ 0.5 eV for the charge flow. It is found that a thickness of  1 u.c. of GTO would be sufficient to allow for the band bending and conduction. Our study shows the importance of correlation in tuning electronic structure of both the interface and the surface layers. Although GGA+$U$ has been successful in this case to correctly capture the correlation effects, we shall demonstrate in the next example that in several cases GGA+$U$ is not sufficient to correctly describe electronic behaviour in correlated systems.

Among the various  novel properties of heterointerfaces, a lot of effort has been made to control and utilize magnetic properties as well for e.g. magneto-electric coupling, magnetic ordering modification, and charge-transfer effect were observed by using manganese oxides as a constituent layer within oxide heterostructures. In this context, LaMnO$_3$(LMO), a Jahn Teller driven insulator, based layers have been widely adopted to employ magnetism to the oxide heterostructures. The primary advantage of using LMO is that its magnetic and electric phases can be modified diversely by a small amount of doping. Stoichiometric LMO is an A-type antiferromagnet and is a good insulator. The system can be doped by cautiously controlling its stoichiometry and can become a ferromagnetic metal. On the other hand, such a doping effect can also be a major disadvantage in identifying a system. The delicate effect of doping requires the careful characterization of LMO, especially when it is used in a heterostructure.

A lot of experiments have been done to identify the nature of the LMO/STO interfaces and the varied nature of the electronic structure of the LMO/STO interfaces have been reported ranging from ferromagnetic metal, antiferrogmanetic insulator, ferromagnetic insulator and even superparamagnetism, depending on the relative thickness of LMO and STO and their geometry \cite{barriocanal-ch6}, \cite{garcia-ch6}, \cite{choi-ch6}, \cite{liu2-ch6}, \cite{sumilan-ch6}. 

Among the several experimental studies that have been undertaken to identify the nature of the LMO/STO interfaces and it was found that the electronic structure of the LMO/STO interface depends strongly on the relative thickness of LMO and STO and in which geometry it is being studied  Having been studied in both superlattice and thin-films geometry, the interfaces  also show strong dependence on the type of geometry in which it is studied. In the superlattice geometry, it is seen that when LMO is much thicker than STO one obtains a ferromagnetic metal, however when LMO and STO have comparable thickness one obtains experimentally a ferromagnetic insulator. There is however no consistent satisfactory explanation of this FM insulating state! In the thin-film/substrate geometry for thickness of LMO $\leq$5 unit cells,  LMO is anti-ferromagnetic. However when thickness of LMO $\geq$ 6 unit cells, LMO is FM and in these cases FM state is usually accompanied with insulating behaviour above the critical thickness. 

\begin{figure}
	\centering
	\includegraphics[width=0.75\columnwidth]{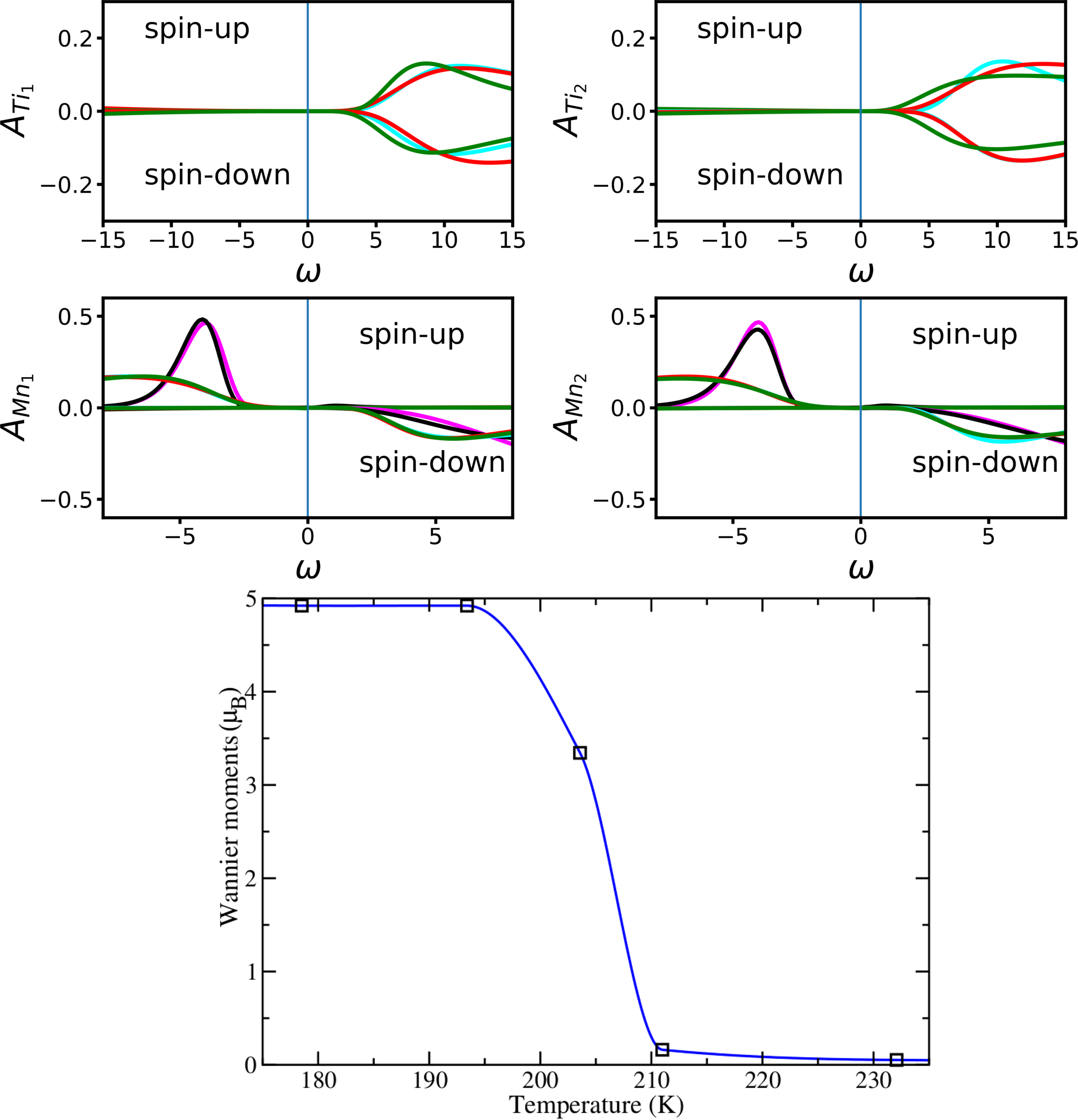}
	\caption{Figure showing the emergence of FM insulating state in LMO/STO from DMFT calculations. The top panel shows the ppectral functions for Ti $t_{2g}$ [$d_{xy}$(cyan), $d_{yz}$(red), $d_{xz}$ (green)] and Mn $t_{2g}$ [color is the same as for Ti] orbitals showing the ferromagnetic insulating phase at $\beta=60$\,eV$^{-1}$. Ti$_1$, Ti$_2$, Mn$_1$, Mn$_2$ belong to different inequivalent layers of the (LMO)$_{3.5}$/(STO)$_{2.5}$ superlattice and form the four impurities of the system. The bottom panel shows the plot of Wannier magnetic moments of Mn as function of temperature, giving the Curie temperature. Figure adapted from Banerjee et al \cite{banerjee-dmft}}
\end{figure}

In our previous studies \cite{banerjee-hse,banerjee-dmft} we show the immense importance of exchange and correlation in case of LMO/STO heterostructures. In these studies we investigated the case of superlattice structure with comparable thicknesses of LMO and STO and we find that we can qualitatively reproduce the experimental result of a ferromagnetic insulating ground state. Within the paradigm of hybrid functionals the microscopic analysis of our results reveals that LMO block becomes insulating owing to the charge disproportionation which is again a purely electronic phenomena as was in the case of epitaxially strained LMO, however the case of STO layers become insulating remained a mystery within this framework, and the fate of the itinerant electron gas generated due to polar catastrophe model and the mechanism driving it insulating was an exciting problem that demanded clarification.
With the aim to provide an understanding for the ferromagnetic insulating state in LMO/STO heterostructures, we applied \textit{ab-initio} DFT+DMFT methods to both superlattices of LMO/STO and “strained-bulk” LMO structures. We investigated the case of LMO/STO superlattice structures with comparable thicknesses of LMO and STO. The ground state was found to be insulating in both paramagnetic and ferromagnetic DFT+DMFT calculations. The results of the spin split DMFT calculations for LMO/STO is shown in Fig. 4. Even when starting from spin polarised DFT calculations and employing a DMFT scheme we find a ferromagnetic insulating state. In all heterostructure geometries that we considered, the 2DEG was found to reside on the LMO side of the interface, contrary to DFT results, as seen from Fig. 4. Though this has been suggested before, this is the first time that an actual first-principles calculation shows the doping of Mn due to the 2DEG. We also showed that the transition temperature from paramagnetic to ferromagnetic phase is high enough to be of practical relevance and accessible to experimental studies, as shown in the bottom panel of Fig. 4.

Another highly interesting case of conductivity and magnetism at oxide heterointerface arises at an interface between LaMnO$_3$(LMO) and SrMnO$_3$(SMO)\cite{salvador-ch5}.
Individually, both these oxides are antiferromagnetic
insulators in their bulk ground state, however at their interface,
a double-exchange ferromagnetism arises in analogy to the behavior of their bulk solid solution, the famous “colossal
magnetoresistance” manganites. Here, the interface charge reconstruction has been experimentally observed by resonant x-ray scattering\cite{smadici-ch5}, providing another intriguing instance of novel two-dimensional states that may be induced by manipulating oxide heterostructures.

Thus we see from our studies that in most oxide heterostructures electronic exchange and correlation plays a very significant role in the correct description of the novel states arising in the 2D electron gas. DMFT calculations turn out to be a very powerful technique in describing such unique electronic states. Prediction of phase transitions alongwith the respective transition temperatures is also possible within the framework of ab-initio DMFT studies. It is thus imperative to properly account for correlations in studies involving complex oxides and its heterostructures, as can be seen from the numerous examples provided in this review.

\section{Outlook}
Although quite a bit of work has been done on oxides not much is known about the full phase diagram of correlated oxides under strain. Some suggestions have been made previously that strong correlations play a significant role in manipulating electronic structure properties of oxides.
As mentioned previously, a relation between strain and strong correlations has been seen from a first principles DFT perspectives in materials like several strongly correlated oxides. IT will be a worthwhile effort to investigate the entire phase space of correlated oxides under strain.

Investigation of structural phase transitions under strain/pressure, structural and magnetic phase transition with a relatively high Curie temperature, the differential effect of tensile and compressive strains provides an interesting playground for strain tuning studies. Similarly the effect of magnetic anisotropy and correlations may also be studied as a function of strain, in spin orbit coupled oxides. Unconventional metal insulator phase transitions depending on type of strain applied along with the related negative magnetoresistance at low temperatures are also worth theoretical investigation.

Several interesting phenomena like temperature dependent correlation driven Mott transitions, high $T_C$ ferromagnetic behaviour in 2D electron gases, topological hall effects coupled with anomalous Hall response, layer- and temperature-dependent multi-orbital metal-insulator transitions in correlated heterostructures have been observed experimentally, albeit without much theoretical understanding. Layer dependence of properties also plays an important role and this may be considered for further investigations as well.

Recently, emergence of interface-stabilized
skyrmions was reported in SRO/SrIrO$_3$ heterostructures via
the measurements of the topological Hall effect in this system.
By investigating the thickness dependence of SRO/SIO bilayer, it was suggested that the skyrmion phase in this bilayer is driven by strong spin-orbit coupling of SIO , which
in combination with octahedral distortion leads to a sizeable
Dzyaloshinskii-Moriya interaction at the interface thus lead-
ing to the formation of chiral structures. Correlation driven spin orbit coupled phases in SIO/STO, as well as SRO/SIO may be looked at in much detail since both involve a very interesting interplay of spin orbit coupling and strong correlations.

Another interesting aspect of theoretical investigation of oxide interfaces may be exotic heterostructure geometries of oxides, some of which has been studied experimentally and exotic phase transitions have been found without much theoretical insight while some of which may in future be studied experimentally. Heterostructures also pave the way for studies on interesting quantum confinement problems related to correlations. 

Huge number of applications have been predicted for interfaces of these complex oxides. Some of these applications have been included in field-effect devices, sensors, photodetectors, and thermoelectrics and in functional solar cells. Thus oxide electronics is thought to one day replace for the better the silicon semiconductor industry as we know it today. Just as Nobel laureate Herbert Kroemer mentioned in his Nobel lecture "the interface is the device"\cite{nobel-ch1}, it indeed seems to be the way to look to the future.

\section*{Acknowledgments}
The author thanks Prof. Markus Aichhorn for productive and useful discussions on the topic. The author is currently funded by the Austrian Science Fund (FWF), START project Y746.

\end{document}